\documentclass[aps,pra,twocolumn,groupedaddress,superscriptaddress,showpacs]{revtex4}
\usepackage{natbib}
\usepackage[latin1]{inputenc}
\usepackage[dvips]{graphicx}
\usepackage{amsmath,amsbsy,amsfonts,amssymb}
\usepackage{bm}
\usepackage{textcomp}

\begin{document}
\title{The $^1$S$+^1$S asymptote of Sr$_2$ studied by Fourier-transform spectroscopy}
\author{Alexander Stein}
\affiliation{Institut f\"ur Quantenoptik, Leibniz Universit\"at Hannover,
Welfengarten 1, D-30167 Hannover, Germany}
\author{Horst Kn\"ockel}
\affiliation{Institut f\"ur Quantenoptik, Leibniz Universit\"at Hannover,
Welfengarten 1, D-30167 Hannover, Germany}
\author{Eberhard Tiemann}
\affiliation{Institut f\"ur Quantenoptik, Leibniz Universit\"at Hannover,
Welfengarten 1, D-30167 Hannover, Germany}
\date{\today}
\begin{abstract}
An experimental study of the long range behavior of the ground state X$^1\Sigma^+_g$ of Sr$_2$ is performed by high resolution spectroscopy of asymptotic vibrational levels and the use of available photoassociation data. Ground state levels as high as v$''=60$ (outer turning point at 23 \AA{} and 0.1 cm$^{-1}$ below the asymptote) could be observed by Fourier-transform spectroscopy of fluorescence progressions induced by single frequency laser excitation of the v$'=4$, J$'=9$ rovibrational level of the state 2$^1\Sigma^+_u$. A precise value of the scattering length for the isotopologue $^{88}$Sr$_2$ is derived and transferred to all other isotopic combinations by mass scaling with the given potential. The derived potential together with already published information about the state 2$^1\Sigma^+_u$ directs to promising optical paths for producing cold molecules in the electronic ground state from an ultracold ensemble of Sr atoms.
\end{abstract}
\pacs{34.20.Cf, 31.50.Bc, 33.20.Kf, 33.20.Vq} 
\keywords{Interatomic potentials and forces, potential energy surfaces for ground electronic states, visible spectra, vibration--rotation analysis}
\maketitle 

\section{Introduction}
\label{intro}

Our spectroscopic work on this molecule is motivated by the current high interest in ultracold ensembles of Strontium atoms \cite{Mickelson,FerrariBO2006,Daley,Takamoto}, which could be a candidate for an optical frequency standard \cite{Boyd2007,Blatt,Ludlow2008}, and also by interest in ultracold Sr$_2$ molecules \cite{Zelevinsky2008,ZelevinskiC2008}. Though for all these experiments reliable knowledge of collision properties for Sr atoms like scattering lengths would be of advantage, there was no sufficiently precise ground state potential available from \textit{ab initio} calculations or experimental work, from which these quantities could be derived by directly solving the radial Schr\"o\-din\-ger equation. 

In \cite{SteinSr2008} we reported on the optical spectrum of the system 2$^1\Sigma^+_u$ --- X$^1\Sigma^+_g$ with more than 10300 transitions which involve the term energies of nearly 60\% of the existing rovibrational levels of the ground state for the main isotopologue $^{88}$Sr$_2$ and additionally many levels for the isotopologues $^{86}$Sr$^{88}$Sr and $^{87}$Sr$^{88}$Sr. We reached vibrational levels up to v$''=49$ and rotational levels up to J$''=224$. Using this data set we had to correct the rotational assignment given by \cite{Gerber_Sr2_1984} by four units and were able to construct a potential energy curve (PEC) which is reliable in the range from 4 to 11 \AA{}, covering an energy region from the bottom up to 9 cm$^{-1}$ below the dissociation asymptote. With the help of theoretically calculated long range coefficients C$_6$ and C$_8$ \cite{Mitroy2003,Porsev2006} and at that time available node positions of the scattering wave function at large internuclear separation for a kinetic energy of a few microkelvins \cite{Yasuda} and 2 mK \cite{Mickelson} from photoassociation, we estimated a complete set of scattering lengths for all combinations of naturally abundant Sr isotopes. Using our derived potentials to calculate Franck-Condon factors we proposed spectroscopic schemes to measure the highest existing vibrational levels of the ground state. Meanwhile new work on photoassociation \cite{MartinezDeEscobar} was performed and the binding energy of the last vibrational level v$''=62$ for $^{88}$Sr$_2$ was determined and for $^{84}$Sr two groups reached Bose-Einstein condensation \cite{BEC1,BEC2}.

In this work we report on successful experiments in the proposed direction and derive ground state levels close to the asymptote with which we extend the precisely known potential region significantly into the long range regime. With these data the experimental knowledge about the long range coefficients C$_6$, C$_8$ and C$_{10}$ and the precision in the scattering lengths will be improved. 

The paper is organized as follows: Section \ref{sec:exp} describes the experimental methods, Sec. \ref{sec:analysis} shows the obtained data set and discusses the measurement uncertainties, Sec. \ref{sec:results} presents the resulting ground state potential and scattering lengths. Sec. \ref{sec:conclusion} gives conclusions and a short outlook.

\section{Experiment}
\label{sec:exp}

The apparatus is the same as in \cite{SteinSr2008}, the experimental procedure is an extension of the earlier one. Strontium is filled into a stainless steel heatpipe, which is heated to a temperature of about 1220 K under 20 mbar of argon as a buffer gas. To keep the optical path free from condensing strontium crystals growing at both ends of the heatpipe, the oven has to be moved every two hours to bring this Sr back to the heated zone. This means, that the heated cell is not working perfectly as a heatpipe where the condensed substance flows back to the heated region.

The molecules are excited to the v$'=4$ level of the state 2$^1\Sigma^+_u$ by light from a Coherent CR 699 ring dye laser operated with Rhodamine 6G at frequencies close to 17635 cm$^{-1}$ (depending on the rotational quantum number J). The power of the dye laser was about 70 mW in front of the heatpipe. The fluorescence light emitted antiparallel to the beam direction of the laser is imaged into a Bruker IFS 120 HR Fourier-transform spectrometer.

To measure the highest vibrational levels of the ground state, the vibrational band of the excited state with v$'=4$ and the rotational levels with J$'=25$, J$'=21$, J$'=17$, J$'=13$, and J$'=9$ were selected. Low J are very important for observing asymptotic levels because high J are not existing for these v. For measuring the progression starting from J$'=25$ a resolution of 0.05 cm$^{-1}$ was used, but because of the small spacings between the lines with  v$''>50$ it was expected that for the smaller J$'$ the P, R doublets of the highest vibrational levels would not be completely resolved with this resolution. Therefore, the resolution was increased for the remaining measurements to 0.02 cm$^{-1}$. For increasing the sensitivity of the measurements without saturating the amplifiers in the detection unit of the spectrometer selected optical filters were used. For the progressions of the four larger J$'$ (25 to 13) a holographic notch filter and a OG 590 color glass filter were inserted to suppress the internal HeNe laser of the spectrometer and the stray light of the dye laser, respectively. For the three progressions (J$'$ = 25, 21, and 17) up to 600 scans of the spectrometer were averaged and for J$'=13$ about 800 scans were necessary to identify safely lines to the highest existing levels with v$''=59$ and 60. To be able to observe v$''=60$ also for J$'=9$, where the intensity is already significantly lower because of the smaller degeneracy of that level, the experiment had to be further optimized. With a single bandpass filter with 10 nm FWHM (\underline{f}ull \underline{w}idth at \underline{h}alf \underline{m}aximum) around 600 nm, where the expected fluorescence for v$'' > 30$ should appear, and a laser power of more than 200 mW in front of the heatpipe it was possible to identify two more rotational levels for v$''=60$ (J$''=8$ and J$''=10$) by averaging 600 scans.

The method, mentioned in \cite{SteinSr2008} and successfully used  for Ca$_2$ in \cite{Allard_2003}, of excitation spectroscopy by tuning the laser over the transition frequencies to the desired levels and filtering the detected fluorescence with a monochromator, was not used for observing even higher v$''$, i.e. 61 and 62, because for these levels, which exist only for J$''\leq8$, the line separations of the P, R doublets become already smaller than the Doppler widths of the individual lines. Beam experiments would be needed to obtain sufficient resolution for this purpose.

\section{Data set}
\label{sec:analysis}

\begin{figure}
\resizebox{0.5\textwidth}{!}{%
\includegraphics{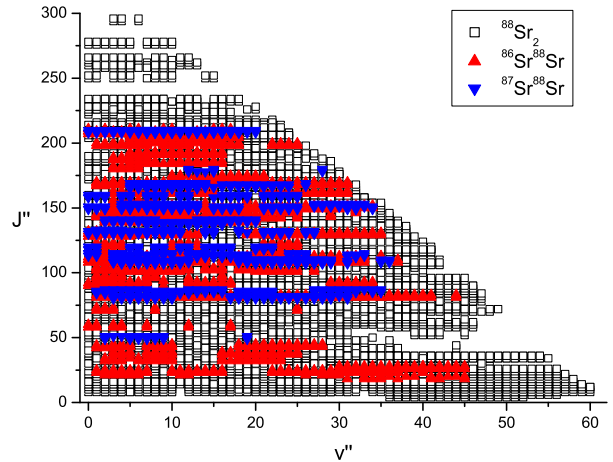}}
\caption{(Color online) Overview of the observed energy levels of the ground state X$^1\Sigma^+_g$, including all measurements of ref. \cite{SteinSr2008}.}
\label{XData}
\end{figure}

The data set published in \cite{SteinSr2008} consisted of 4673 different rovibrational levels for the ground state, containing 3163 of the ca. 5400 existing ground state levels of the main isotopologue $^{88}$Sr$_2$. The current ground state data set is extended to 6077 different rovibrational levels, of which 4159 belong to the isotopologue $^{88}$Sr$_2$ and 1301 and 617 belong to the isotopologues $^{86}$Sr$^{88}$Sr and $^{87}$Sr$^{88}$Sr, respectively. The complete data field is shown in figure \ref{XData}.

The assignment of the 15700 lines (5400 new since publication of \cite{SteinSr2008}) of the current data set was done using a partially automated software as already described in \cite{SteinSr2008} (in a prior version in \cite{SteinLiCs2008}). This software uses the differences between individual lines of a selected progression to automatically assign them to a molecule and an electronic transition and to quantum numbers v$'$, J$'$, v$''$ and J$''$, provided that the data of the ground and as much as available of the excited states of the molecules are input to the program. Additional lines which cannot be automatically identified to belong to the progression, because they extend significantly beyond the precisely known region of the ground state potential, are individually selected by the user before the program automatically assigns the best fitting quantum numbers.

In \cite{SteinSr2008} the uncertainties for the differences of the line frequencies were estimated to be 0.005 cm$^{-1}$ in most cases. This is one tenth of the applied resolution of 0.05 cm$^{-1}$. We keep this uncertainty for the lines which are taken from the earlier data set and also the lines newly measured using the same resolution. For the new lines measured with the higher resolution of 0.02 cm$^{-1}$ we estimate a smaller uncertainty of 0.002 cm$^{-1}$ for the stronger lines (S/N ratio larger than 5). For these lines a significant contribution of any Doppler shift can be excluded, since the excitation frequency was optimized on the center of the Doppler profile and thus the velocity components of the excited molecules parallel to the direction of observation will be very small. The specified uncertainty of the Fourier spectrometer for frequency differences within a spectrum is 0.001 cm$^{-1}$, still below the assumed experimental uncertainty. For weak lines and lines belonging to accidentally overlapping excitations an uncertainty of 0.004 cm$^{-1}$ is assumed. 

The complete list of assigned lines can be found in the additional online material.

\section{Results}
\label{sec:results}

For the ground state a potential energy curve is calculated using the same type of representation as in \cite{SteinSr2008} (details can be found in \cite{Allard_Ca2_2002}). The central part of the potential ($R_i\leq R\leq R_a$) is described by 

\begin{equation}
\label{GVC}
V_c(R)=T_m+\sum_{j\geq1}a_jx^j
\end{equation} 

\noindent with the nonlinear mapping function

\begin{equation}
\label{GVx}
x=\frac{R-R_m}{R+bR_m} \mbox{ .}
\end{equation}

The inner repulsive wall ($R < R_i$) is given by

\begin{equation}
\label{GVi}
V_i(R)=A+\frac{B}{R^n} \mbox{ ,}
\end{equation}

while the long range part ($R > R_a$) is represented by

\begin{equation}
\label{GVa}
V_a(R)=U_\infty-\frac{C_6}{R^6}-\frac{C_8}{R^8}-\frac{C_{10}}{R^{10}} \mbox{ .}
\end{equation}

The coefficients $a_j$ are the main fitting parameters, while the parameter $T_m$ is adjusted to obtain at the point $R_a$ a continuous connection between the central and the long range part of the potential. $R_m$ in eqn. (\ref{GVx}) is chosen close to the minimum of the potential. The parameters b and $R_i$ are manually chosen to optimize the fitting result with a small number of parameters $a_j$. The real parameter n \cite{n_s} is an additional fitting parameter, while the parameters A and B are adjusted for a continuously differentiable connection at the point $R_i$. 

The radius $R_a$ should be close to the LeRoy radius $R_b(AB)=2\left[\sqrt{\left<r_A^2\right>}+\sqrt{\left<r_B^2\right>}\right]$\cite{LeRoy1974}, where $r_A$ and $r_B$ are the radial coordinates of the outermost electrons of atom $A$ and atom $B$, here $r_A=r_B=r_{Sr}$. The resulting LeRoy radius $R_b$ is 9.7 \AA{} or 10.5 \AA{} for the X state of Sr$_2$, if $\left<r_{Sr}^2\right>$ is taken from \cite{Lu1971} or from \cite{Fischer1972}, respectively. The larger value $R_a=10.5$ \AA{} was chosen for the final potential evaluation. This choice should justify the neglect of a formal term for an exchange energy in the long range part according to eq. (\ref{GVa}).

\begin{figure}
\resizebox{0.5\textwidth}{!}{%
\includegraphics{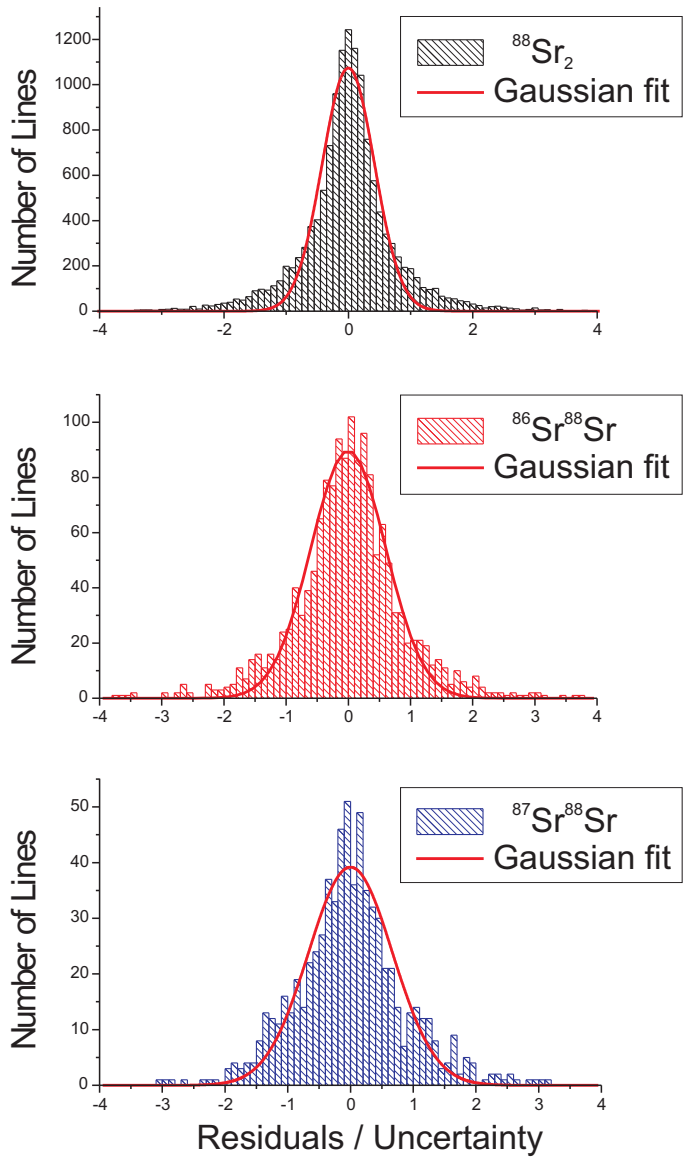}}
\caption{(Color online) The residuals of the X$^1\Sigma^+_g$ potential fit divided by the specific uncertainties represented by histograms compared to Gaussian fits.}
\label{XResiduals}
\end{figure}

The residuals of the potential fit can be found as histograms in figure \ref{XResiduals} compared to Gaussian profile fits. The plot for the main isotopologue $^{88}$Sr$_2$ shows a symmetric profile with a FWHM of 0.8, while the shapes for the other two isotopologues are significantly wider and are not as symmetric as for the main isotopologue. This can be attributed to the much lower number of data for the isotopologues $^{86}$Sr$^{88}$Sr and $^{87}$Sr$^{88}$Sr and on average significantly weaker line intensities, which lead to a less precise determination of the frequencies. 

\begin{table}
\caption{The potential coefficients and derived molecular constants for the ground state X$^1\Sigma^+_g$. The potential energy is calculated with respect to the asymptote, the dispersion coefficients are the recommended values from the third column of table \ref{XPotC}.}
\label{XPotential}
\centering
\begin{tabular}{lp{0.1\textwidth}r}\hline\noalign{\smallskip}
a$_{1}$   && -6.50$\times 10^{-2}$ cm$^{-1}$ \\
a$_{2}$   && 1.5939056$\times 10^{4}$ cm$^{-1}$ \\
a$_{3}$   && -2.9646778$\times 10^{4}$ cm$^{-1}$ \\
a$_{4}$   && -6.269777$\times 10^{3}$ cm$^{-1}$ \\
a$_{5}$   && 4.4952358$\times 10^{4}$ cm$^{-1}$ \\
a$_{6}$   && 8.709016$\times 10^{3}$ cm$^{-1}$ \\
a$_{7}$   && -1.0054929$\times 10^{5}$ cm$^{-1}$ \\
a$_{8}$   && 5.94784152$\times 10^{5}$ cm$^{-1}$ \\
a$_{9}$   && -9.95239126$\times 10^{5}$ cm$^{-1}$ \\
a$_{10}$   && -1.14496717$\times 10^{7}$ cm$^{-1}$ \\
a$_{11}$   && 4.606463055$\times 10^{7}$ cm$^{-1}$ \\
a$_{12}$   && 3.74666573$\times 10^{7}$ cm$^{-1}$ \\
a$_{13}$   && -5.439157146$\times 10^{8}$ cm$^{-1}$ \\
a$_{14}$   && 9.364833940$\times 10^{8}$ cm$^{-1}$ \\
a$_{15}$   && 1.387879748$\times 10^{9}$ cm$^{-1}$ \\
a$_{16}$   && -8.4009054730$\times 10^{9}$ cm$^{-1}$ \\
a$_{17}$   && 1.5781752106$\times 10^{10}$ cm$^{-1}$ \\
a$_{18}$   && -1.5721037673$\times 10^{10}$ cm$^{-1}$ \\
a$_{19}$   && 8.376043061$\times 10^{9}$ cm$^{-1}$ \\
a$_{20}$   && -1.88984880$\times 10^{9}$ cm$^{-1}$ \\
\hline
b     && -0.17    \\
R$_m$ && 4.6719018 \AA \\ 
T$_m$ && -1081.6384 cm$^{-1}$ \\ \hline
R$_i$ && 3.963 \AA \\
n     && 12.362    \\
A     && -1.3328825$\times 10^{3}$ cm$^{-1}$ \\
B     && 3.321662099$\times 10^{10}$ cm$^{-1}$\AA$^n$ \\ \hline
R$_a$ && 10.5 \AA \\
C$_{6}$ && 1.525$\times 10^{7}$ cm$^{-1}$\AA$^{6}$ \\
C$_{8}$ && 5.159$\times 10^{8}$ cm$^{-1}$\AA$^{8}$ \\
C$_{10}$ && 1.91$\times 10^{10}$ cm$^{-1}$\AA$^{10}$ \\ \hline \hline
\multicolumn{2}{l}{derived constants:}       \\
D$_e$    && 1081.64(2) cm$^{-1}$ \\
D$_0$    && 1061.58(1) cm$^{-1}$ \\
R$_e$    && 4.6720(1) \AA \\ \hline
\end{tabular}
\end{table}

The derived potential coefficients are given in table \ref{XPotential}. The normalized standard deviation of this potential fit is $\sigma=0.80$ which indicates that the applied measurement uncertainties are as a whole not underestimated. It correlates to the width of the histogram in fig. \ref{XResiduals} which also gives a normalized FWHM of 0.8 for the main isotopologue. The values of the dissociation energy $D_e$ and the equilibrium internuclear distance $R_e$ depend on the chosen potential description, namely on the number of potential coefficients $a_i$, on the parameter $b$ and on the connection radii $R_i$ and $R_a$. The given uncertainties are 1$\sigma$ standard deviations derived by comparing 35 different potential versions with different parameter settings which all had almost the same $\sigma$ values. Compared to $D_e$ the value for $D_0$ (dissociation energy with respect to (v=0,J=0) for the main isotopologue $^{88}$Sr$_2$) is less model dependent.

\begin{table*}
\caption{Comparison of the long range coefficients derived from fits of freely varying parameters with the recommended values applying a fixed $C_8$ value and with theoretical long range coefficients published in recent years. For the details on evaluation procedure and uncertainty estimates see text.}
\label{XPotC}
\centering
\begin{tabular}{lccccccc}\hline\noalign{\smallskip}
 & freely var. & recommended & \cite{Mitroy2003} & \cite{Porsev2006} & \cite{Stanton1994} & \cite{Porsev2002} & \cite{Lima2005} \\ \hline\noalign{\smallskip}
$C_6$  [$10^7 $cm$^{-1}$\AA$^6$]       & 1.527(33)    & 1.525(5)          & 1.566 & 1.495(3)   & 1.5481 & 1.528(94) & 1.509(20) \\
$C_8$  [$10^8 $cm$^{-1}$\AA$^8$]       & 5.1(15)     & (5.159)\cite{aC8} & 5.201 & 5.118(4)   &        &           &           \\
$C_{10}$ [$10^{10}$cm$^{-1}$\AA$^{10}$]& 2.0(13)     & 1.91(27)          & 1.61  & 1.593      &        &           &           \\ \hline\
\end{tabular}
\end{table*}

For the first potential fits the long range coefficients $C_6$, $C_8$, and $C_{10}$ were treated as free parameters. Despite the precision of the energies of the levels v$''=62$, J$''=0$ and J$''=2$ from \cite{MartinezDeEscobar} incorporated in the fit, the set of coefficients is still significantly correlated. While the $C_6$ coefficient is quite well fixed the coefficients $C_8$ or $C_{10}$ can be varied over a larger range of values without a significant reduction in the quality of the description of the data set, if all other coefficients are adjusted accordingly. To estimate uncertainties for the long range coefficients different potential fits were done where each time the $C_8$ coefficient was held fixed at a different value within its range as given in the second column of table \ref{XPotC}. The results shown for the other $C_i$ coefficients are averaged values from the series of fits and their uncertainty bounds represent an increase of the sum of the weighted squared residuals by 2\%, which is a quite safe limit, since only 113 of the 6077 energy levels included in the fit are directly affected by this change.

Because the derived uncertainty for the $C_8$ coefficient is significantly larger than the uncertainty claimed for the theoretically calculated coefficient by \cite{Porsev2006} (see the fifth column of table \ref{XPotC}) and also larger than the difference between the two available theoretical $C_8$ coefficients \cite{Mitroy2003,Porsev2006}, additional potential fits were done, where each time the $C_8$ coefficient was held fixed at a value taken from \cite{Mitroy2003} or \cite{Porsev2006} or the average of both. Such a procedure was also tried with the theoretical $C_6$ coefficients, but those resulting potentials gave a significantly less convincing description of the data set and also the resulting coefficients $C_8$ and $C_{10}$ showed strong deviations to the theoretical values. Additional fits were done for different numbers of potential coefficients $a_j$ and different connection radii $R_i$ and $R_a$. For a few fits $R_a$ was set to values as large as 13 \AA{}, which led to significantly less well determined long range coefficients with higher correlation without obtaining a significant improvement in the description of the data. The resulting $C_i$ coefficients of all the fits using either the $C_8$ coefficient from \cite{Mitroy2003} or from \cite{Porsev2006} or its mean value were averaged to obtain the recommended values, which are listed in the third column of the table \ref{XPotC}. The uncertainties are 1$\sigma$ standard deviations of the averaged values.

Table \ref{XPotC} compares the dispersion parameters derived with freely varied $C_i$ coefficients and the recommended values with the published theoretical long range coefficients from recent years. As one can see, the values in the second column are in good agreement with all the other values, while the recommended values do deviate from the most precise theoretical coefficient set from \cite{Porsev2006} more than one would expect considering the given uncertainties, there. This could be due to different reasons. The paper \cite{Porsev2006} used the lifetime of the atomic $^1$S---$^1$P transition from \cite{Yasuda}, which was calculated from the $C_3$ coefficient of the state 2$^1\Sigma^+_u$ fitted to photoassociation data. In \cite{Yasuda} line broadenings of the PA resonances are mentioned, which could indicate predissociation due to the coupling to the continuum of electronic states from lower asymptotes and could lead to a systematic shift of these asymptotic levels of the state 2$^1\Sigma^+_u$. Another reason could be that the presently derived uncertainties are underestimated because of a too low number of different potential descriptions used for averaging. The computing effort of testing the long range correlation by a Monte Carlo study would be enormous for the presently employed analytical type of potential description and the available hard and software.

A pointwise representation of the potential derived from the parameters in table \ref{XPotential} and the analytical potential description for the best fitting values obtained by freely varying $C_i$ coefficients can be found in the additional online material.

For the convenience of the reader we also made fits using the MLR potential description from \cite{LeRoy_pre}, which might allow for defining the potential by a lower number of parameters. It has the mathematical form

\begin{equation}
\label{MLR}
V_{MLR}(R)=D_e\left[1-\frac{U_{LR}(R)}{U_{LR}(R_e)}e^{-\beta(y^{ref}_p)\times{}y^e_p(R)}\right]^2
\end{equation}

with the long-range function

\begin{equation}
\label{uLR}
U_{LR}(R)=\frac{C_6}{R^6}+\frac{C_8}{R^8}+\frac{C_{10}}{R^{10}} 
\end{equation}

and the exponential function

\begin{equation}
\label{beta}
\beta =y^{ref}_p(R)\beta_\infty + \left[1-y^{ref}_p(R)\right]\sum^N_{i=0}\beta_i y^{ref}_q(r)^i \mbox{ ,}
\end{equation}

where 

\begin{equation}
\label{betaInf}
\beta_\infty=\ln\left[\frac{2D_e}{U_{LR}(R_e)}\right] \mbox{ .}
\end{equation}

Three mapping functions $y^a_b$ are defined as

\begin{equation}
\label{map}
y^a_b=\frac{R^b-R^b_a}{R^b+R^b_a} \mbox{ .}
\end{equation}

where the placeholder $a$ can either be '$e$' for the equilibrium radius $R_e$ or '$ref$' for $R_{ref}$ and the exponent $b$ can either be $p$ or $q$. $R_{ref}$ is an expansion point at an internuclear distance larger than $R_e$. The $\beta_i$, $C_i$, $R_e$, and the dissociation energy $D_e$ are the fit parameters, while the parameter $R_{ref}$ and the exponents $p$ and $q$ are manually chosen to optimize the potential description with a minimum number of coefficients and to get the proper long range behavior.

\begin{table}
\caption{The coefficients of the MLR potential description for the ground state X$^1\Sigma^+_g$. The potential energy is calculated with respect to the potential minimum, the dispersion coefficients were taken from table \ref{XPotC}.}
\label{XPotMLR}
\centering
\begin{tabular}{lp{0.1\textwidth}r}\hline\noalign{\smallskip}
$\beta_{0}$   &&  -1.458760592449 \\
$\beta_{1}$   &&  -0.031511799120 \\
$\beta_{2}$   &&  -0.912033020316 \\
$\beta_{3}$   &&  -0.033913017557 \\
$\beta_{4}$   &&  -0.465124882150 \\
$\beta_{5}$   &&   0.287008644909 \\
$\beta_{6}$   &&  -0.585285048765 \\
$\beta_{7}$   &&   2.490073196261 \\
$\beta_{8}$   &&   3.065232204927 \\
$\beta_{9}$   &&  -8.521057394373 \\
$\beta_{10}$  &&  -0.771989570637 \\
$\beta_{11}$  &&  17.679758812808 \\
$\beta_{12}$  && -11.143746759192 \\
$\beta_{13}$  &&  -1.704727367184 \\
$C_{6}$       && 1.525$\times 10^{7}$ cm$^{-1}$\AA$^{6}$ \\
$C_{8}$       && 5.159$\times 10^{8}$ cm$^{-1}$\AA$^{8}$ \\
$C_{10}$      && 1.910$\times 10^{10}$ cm$^{-1}$\AA$^{10}$ \\
$D_e$         && 1081.6352 cm$^{-1}$ \\
$R_e$         && 4.6719032 \AA \\
$R_{ref}$     && 5.5 \AA    \\
$p$           && 5 \\ 
$q$           && 3 \\ \hline
\end{tabular}
\end{table}

The potential coefficients for the MLR description are given in table \ref{XPotMLR}, the normalized standard deviation is $\sigma=0.80$ as for the description given in table \ref{XPotential}. Thus there is no significant difference in the quality of the description of the data set. The clear advantage of the potential description from table \ref{XPotMLR} compared to the potential description from table \ref{XPotential} is the lower number of potential coefficients necessary (14 $\beta_i$ instead of 20 $a_i$) and the fact that it is indefinitely continuously differentiable at all points which is not the case for the potential of table \ref{XPotential} at the points $R_i$ and $R_a$. Disadvantages are the much more complex formula with three mapping functions instead of one and a much stronger correlation of the long range coefficients $C_i$ with the potential coefficients $\beta_i$ (see eqn (\ref{MLR})). For the potential description from table \ref{XPotential} the correlation between the $C_i$ and the $a_i$ is only due to the integration of the wave functions over the full potential region. Thus we applied the recommended $C_i$ values from table \ref{XPotC} for the fit of the MLR potential. The dissociation energy obtained in the MLR approach agrees very well to the derived value in table \ref{XPotential}.

\begin{table*}
\caption{The scattering lengths in units of $a_0$ (Bohr radius) derived with different long range estimations and compared to earlier values published in 2008. The hyperfine splitting of the isotope $^{87}$Sr is neglected.}
\label{Scat}
\centering
\begin{tabular}{rcccc}
\hline\noalign{\smallskip}
isotopes            & $C_i$ freely var. & recommended $C_i$ & \cite{MartinezDeEscobar} & \cite{SteinSr2008} \\ \hline\noalign{\smallskip}
$^{84}$Sr+$^{84}$Sr & 122.58(72) & 122.762(92) & 122.7(3)       & 121 to 127  \\
$^{84}$Sr+$^{86}$Sr & 31.4(11)   & 31.65(14)   & 31.9(3)        & 30 to 36    \\
$^{84}$Sr+$^{87}$Sr & -58.6(43)  & -57.61(61)  & -56(1)         & -64 to -45  \\
$^{84}$Sr+$^{88}$Sr & 1561(400)  & 1658(54)    & 1790(130)      & $\geq$1170 or $\leq$-1900 \\
$^{86}$Sr+$^{86}$Sr & 777(84)    & 798(12)     & 823(24)        & 677 to 1430 \\
$^{86}$Sr+$^{87}$Sr & 161.9(15)  & 162.25(21)  & 162.5(5)       & 160 to 171  \\
$^{86}$Sr+$^{88}$Sr & 97.24(53)  & 97.374(69)  & 97.4(1)        & 97 to 101   \\
$^{87}$Sr+$^{87}$Sr & 96.06(53)  & 96.198(68)  & 96.2(1)        & 95 to 99    \\
$^{87}$Sr+$^{88}$Sr & 54.64(73)  & 54.819(92)  & 55.0(2)        & 54 to 58    \\
$^{88}$Sr+$^{88}$Sr & -2.5(20)   & -2.00(27)   & -1.4(6)        & -4.8 to 4.5 \\
\hline
\end{tabular}
\end{table*} 

Table \ref{Scat} shows the derived scattering lengths of the different isotopic compositions and compares them to other recent publications. The values in the second column were calculated from the potentials which were applied for the estimation of the uncertainty of the long range coefficients as given in the second column of table \ref{XPotC}. The values in the third column of table \ref{Scat} are averages from different potentials which were obtained using $C_8$ coefficients from \cite{Mitroy2003} and \cite{Porsev2006} and resulted in the recommended values in the third column of table \ref{XPotC}. It turned out that the influences of different connection radii $R_i$ or $R_a$ or numbers of parameters $a_j$, which causes mainly a change in the central part of the potential, has a much larger influence on the scattering lengths than the question if the $C_8$ should be taken from \cite{Mitroy2003} or from \cite{Porsev2006}. All of the potentials used reproduce the measured data nearly equally well and also the node positions of the scattering wave functions from photoassociation \cite{Mickelson,Yasuda}. The uncertainties given in the second and third column of table \ref{Scat} are 1$\sigma$ standard deviations in units of the last digit shown. The uncertainties in the fourth column are taken from \cite{MartinezDeEscobar}. The last column provides the results from our earlier analysis in \cite{SteinSr2008} for comparison. The very precise analysis of the present work is consistent with the values from \cite{SteinSr2008} and with those from the photoassociation work \cite{MartinezDeEscobar}. The new determination yields about a factor of 2 smaller uncertainties than reported in \cite{MartinezDeEscobar}.

\section{Conclusion}
\label{sec:conclusion}

After publishing a description for the ground state potential for the region from 4 to 11 \AA{} and the vibrational interval from v$''=0$ to 48 in \cite{SteinSr2008}, for the present paper we concentrated our work on the asymptotic levels of this state. By applying the same spectroscopic method but improving the detection scheme, increasing the signal averaging time and selecting appropriate excited rovibrational levels to reach the long range regime we were able to observe levels with v$''$ up to 60, while v$''=62$ would be the last existing one for low J$''$. This allowed us to extend the potential region up to an internuclear distance of 23 \AA~from our data set. 

Including the term energies of the v$''=62$, J$''=0$ and J$''=2$ levels measured by photoassociation \cite{MartinezDeEscobar}, we were able to derive the long range coefficients $C_6$, $C_8$ and $C_{10}$. By taking the $C_8$ coefficient from theory \cite{Mitroy2003,Porsev2006}, which has in our modeling a larger uncertainty than the theoretically calculated $C_8$ coefficient, we derived an internally consistent set of the remaining two coefficients $C_6$ and $C_{10}$. This leads to a further reduction of the uncertainties in the calculation of the scattering lengths for all combinations of natural abundant isotopes of Strontium. We were also able to give a reliable value for the dissociation energy $D_0$ of the X state. 

The combination of the data of both the state 2$^1\Sigma^+_u$ \cite{SteinSr2008}, and the ground state X$^1\Sigma^+_g$ can be useful for experiments producing ultracold molecules. The transition path successfully applied for this work to observe vibrational levels of the ground state close to the asymptote can be used in the reverse direction to transfer population of highly excited ground state levels to deeply bound vibrational levels of the X state within a single STIRAP step \cite{BergmannSTIRAP}. The asymptotic ground state levels, which can be populated e.g. by spontaneous decay after photoassociation, show the best Franck-Condon overlap with the v$'=4$, v$'=3$ and v$'=9$ levels of the state $2^1\Sigma^+_u$. While either the v$'=4$ or the v$'=3$ level give good overlap with almost every level of the X state, by excitation of the v$'=9$ level almost 12\% of the spontaneous decay reaches the v$''=0$ level. By the use of the $2^1\Sigma^+_u$ potential determined in \cite{SteinSr2008} together with the ground state potential from this work the transition frequencies from all ground state levels to the minimum region (v$'$=0 - 12) of the state $2^1\Sigma^+_u$ should be easily calculable with a precision better than 0.01 cm$^{-1}$. This accuracy should reduce substantially the time for finding STIRAP transitions.

For the near future we plan to further investigate the excited states of the Sr$_2$ molecule. We already started with the state $2^1\Sigma^+_u$ and we will look for the state $1^1\Sigma^+_u$ and a state of type $^1\Pi_u$ for which we found one Q progression during this work.

\section{Acknowledgments}

The authors are grateful to A. Gerdes who helped with the preparation of the dye lasers and to A. Pashov for useful discussions. This work was supported by the Deutsche Forschungsgemeinschaft in the Sonderforschungsbereich 407.

\bibliography{Sr2_Asymptote}

\end{document}